\documentclass[twoside,twocolumn,9pt]{article}
\usepackage{extsizes}
\usepackage[super,sort&compress,comma]{natbib} 
\usepackage[version=3]{mhchem}
\usepackage[left=1.5cm, right=1.5cm, top=1.785cm, bottom=2.0cm]{geometry}
\usepackage{balance}
\usepackage{mathptmx}
\usepackage{sectsty}
\usepackage{graphicx} 
\usepackage{lastpage}
\usepackage[format=plain,justification=justified,singlelinecheck=false,font={stretch=1.125,small,sf},labelfont=bf,labelsep=space]{caption}
\usepackage{float}
\usepackage{fancyhdr}
\usepackage{fnpos}
\usepackage[english]{babel}
\addto{\captionsenglish}{%
  
}
\usepackage{array}
\usepackage{droidsans}
\usepackage{charter}
\usepackage[T1]{fontenc}
\usepackage[usenames,dvipsnames]{xcolor}
\usepackage{setspace}
\usepackage[compact]{titlesec}
\usepackage[hidelinks]{hyperref}

\usepackage{epstopdf}

\definecolor{cream}{RGB}{222,217,201}

\begin{document}

\pagestyle{fancy}
\thispagestyle{plain}
\fancypagestyle{plain}{
\renewcommand{\headrulewidth}{0pt}
}

\makeFNbottom
\makeatletter
\renewcommand\LARGE{\@setfontsize\LARGE{15pt}{17}}
\renewcommand\Large{\@setfontsize\Large{12pt}{14}}
\renewcommand\large{\@setfontsize\large{10pt}{12}}
\renewcommand\footnotesize{\@setfontsize\footnotesize{7pt}{10}}
\makeatother

\renewcommand{\thefootnote}{\fnsymbol{footnote}}
\renewcommand\footnoterule{\vspace*{1pt}%
\color{cream}\hrule width 3.5in height 0.4pt \color{black}\vspace*{5pt}} 
\setcounter{secnumdepth}{5}

\makeatletter 
\renewcommand\@biblabel[1]{#1}            
\renewcommand\@makefntext[1]%
{\noindent\makebox[0pt][r]{\@thefnmark\,}#1}
\makeatother 
\renewcommand{\figurename}{\small{Fig.}~}
\sectionfont{\sffamily\Large}
\subsectionfont{\normalsize}
\subsubsectionfont{\bf}
\setstretch{1.125} 
\setlength{\skip\footins}{0.8cm}
\setlength{\footnotesep}{0.25cm}
\setlength{\jot}{10pt}
\titlespacing*{\section}{0pt}{4pt}{4pt}
\titlespacing*{\subsection}{0pt}{15pt}{1pt}

\fancyfoot{}
\fancyfoot[LO,RE]{\vspace{-7.1pt}\includegraphics[height=9pt]{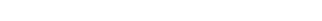}}
\fancyfoot[CO]{\vspace{-7.1pt}\hspace{13.2cm}\includegraphics{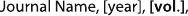}}
\fancyfoot[CE]{\vspace{-7.2pt}\hspace{-14.2cm}\includegraphics{head_foot/RF}}
\fancyfoot[RO]{\footnotesize{\sffamily{1--\pageref{LastPage} ~\textbar  \hspace{2pt}\thepage}}}
\fancyfoot[LE]{\footnotesize{\sffamily{\thepage~\textbar\hspace{3.45cm} 1--\pageref{LastPage}}}}
\fancyhead{}
\renewcommand{\headrulewidth}{0pt} 
\renewcommand{\footrulewidth}{0pt}
\setlength{\arrayrulewidth}{1pt}
\setlength{\columnsep}{6.5mm}
\setlength\bibsep{1pt}

\makeatletter 
\newlength{\figrulesep} 
\setlength{\figrulesep}{0.5\textfloatsep} 

\newcommand{\topfigrule}{\vspace*{-1pt}%
\noindent{\color{cream}\rule[-\figrulesep]{\columnwidth}{1.5pt}} }

\newcommand{\botfigrule}{\vspace*{-2pt}%
\noindent{\color{cream}\rule[\figrulesep]{\columnwidth}{1.5pt}} }

\newcommand{\dblfigrule}{\vspace*{-1pt}%
\noindent{\color{cream}\rule[-\figrulesep]{\textwidth}{1.5pt}} }

\newcommand{\bl}[1]{{\color{black}#1}}
\newcommand{\bll}[1]{{\color{black}#1}}
\newcommand{\ah}[1]{{\color{black}#1}}
\newcommand{\mcm}[1]{{\color{black}#1}}
\newcommand{\mc}[1]{{\color{black}#1}}
\newcommand{\dm}[1]{{\color{black}#1}}

\makeatother

\newcommand{\fig}{Fig.~}
\newcommand{\eq}{Eq.~}

\twocolumn[
  \begin{@twocolumnfalse}
{\includegraphics[height=30pt]
{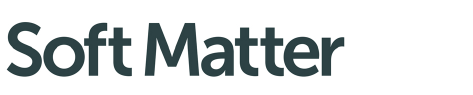}\hfill\raisebox{0pt}[0pt][0pt]
{\includegraphics[height=55pt]{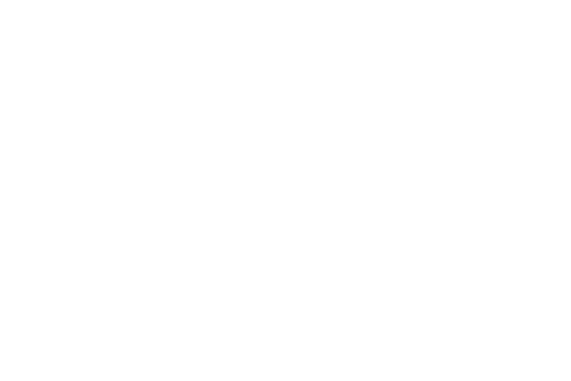}}\\[1ex]
\includegraphics[width=18.5cm]{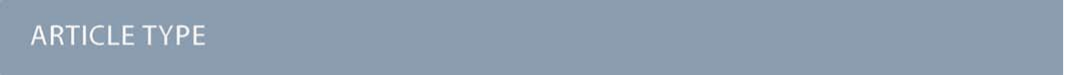}}\par
\vspace{1em}
\sffamily
\begin{tabular}{m{4.5cm} p{13.5cm} }

\includegraphics{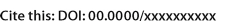} & \noindent\LARGE{\textbf{Motility induced phase separation of deformable cells$^\dag$}} \\
\vspace{0.3cm} & \vspace{0.3cm} \\

 & \noindent\large{Austin Hopkins$^{\ast}$\textit{$^{a}$}, Benjamin Loewe\textit{$^{b\ddag}$}, Michael Chiang\textit{$^{b\ddag}$}, Davide Marenduzzo\textit{$^{b\ddag}$}, and M. Cristina Marchetti$^{\ast}$\textit{$^{a}$}} \\

\includegraphics{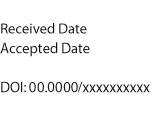} & \noindent\normalsize{Using a multi-phase field model, we examine how particle deformability, which is a proxy for cell stiffness, affects \mcm{motility induced phase separation (MIPS)}. \mcm{We show that purely repulsive deformable, i.e., squishy, cells phase separate more effectively than their rigid counterparts. This} 
can be understood as \mcm{due to the fact that}
deformability \mcm{increases the effective duration of  collisions.}
In addition, the dense regions become increasingly disordered as deformability increases. Our results contextualize the applicability of MIPS to biological systems and have implications for how cells in biological systems may self-organize. } \\

\end{tabular}

 \end{@twocolumnfalse} \vspace{0.6cm}
 
]

\renewcommand*\rmdefault{bch}\normalfont\upshape
\rmfamily
\section*{}
\vspace{-1cm}


\footnotetext{\textit{$^{a}$~Department of Physics, University of California Santa Barbara, Santa Barbara, CA 93106, USA.  E-mail: xxxx@aaa.bbb.ccc}}
\footnotetext{\textit{$^{b}$~School of Physics and Astronomy, University of Edinburgh, Peter Guthrie Tait Road, Edinburgh EH9 3FD, United Kingdom. }}





\section{Introduction}
Self-propelled particles \cite{bechinger2016active} have been used as a simple model for synthetic active swimmers and biological systems, and can describe collective phenomena such as flocking \cite{vicsek1995novel,toner1995long,chate2008collective,cavagna2018physics}, aggregation\cite{buttinoni2013dynamical, yang2014aggregation} and sorting \cite{belmonte2008self, mehes2012collective, mccandlish2012spontaneous}.
Although biological systems often have complex physical interactions, it has been shown that motility is sufficient to induce phase separation of purely repulsive particles.
This phenomenon is known as motility induced phase separation (MIPS) because, unlike in equilibrium systems, the phase separation can occur without attractive interactions.
MIPS has been extensively studied in the context of self-propelled repulsive spheres, known as active Brownian Particles (ABPs), and \mcm{it has} been described in terms of \mcm{the suppression of the} effective motility \mcm{due to crowding}~\cite{tailleur2008statistical, fily2012athermal, farrell2012pattern, bialke2013microscopic, stenhammar2013continuum, fily2014freezing, stenhammar2014phase, cates2015motility}, the kinetics and mechanics of the phase-separated interface 
\cite{redner2013structure,omar2022mechanical}, and an effective attractive interaction \cite{farage2015effective}.
The phase behavior of \mcm{rigid, repulsive active particles} has been mapped \mcm{out} as a function of motility and density \cite{digregorio2018full}, and the effects of other properties like polydispersity \cite{paoluzzi2022motility}, particle shape \cite{grossmann2020particle}, friction between particles \cite{nie2020stability}, and interaction softness \cite{sanoria2021influence,sanoria2022percolation,lauersdorf2021phase} have also been studied.

One shortcoming, however,  is that studies have focused on rigid particles, even though the cells that make up biological systems can change their shape.
Therefore, we study here a system of deformable active particles to better understand the applicability of MIPS to biological systems, or to cell suspensions.
We find that, at a given density, more deformable particles are more prone to phase separate than less deformable ones.
This result can be explained by an increase in the duration of two-body collisions with increasing deformability.
We also find that \mcm{deformability fundamentally affects} the structure of the dense phase, \mcm{which} is crystalline at low deformability and becomes glassy with \mcm{increasing} deformability.

\mcm{In the remainder of the paper we first introduce the phase field model in Section \ref{sec:model}. The results are presented in Section \ref{sec:results}, including the numerically evaluated phase diagram, a phenomenological argument that relates the deformability-induced enhancement of phase separation to the duration of binary collisions, and an analysis the structural properties of the dense phase. We conclude with a brief summary and outlook in Section \ref{sec:summary}.}

\section{Model}
\label{sec:model}

\begin{figure*}
 \centering
 \includegraphics[width=\textwidth]{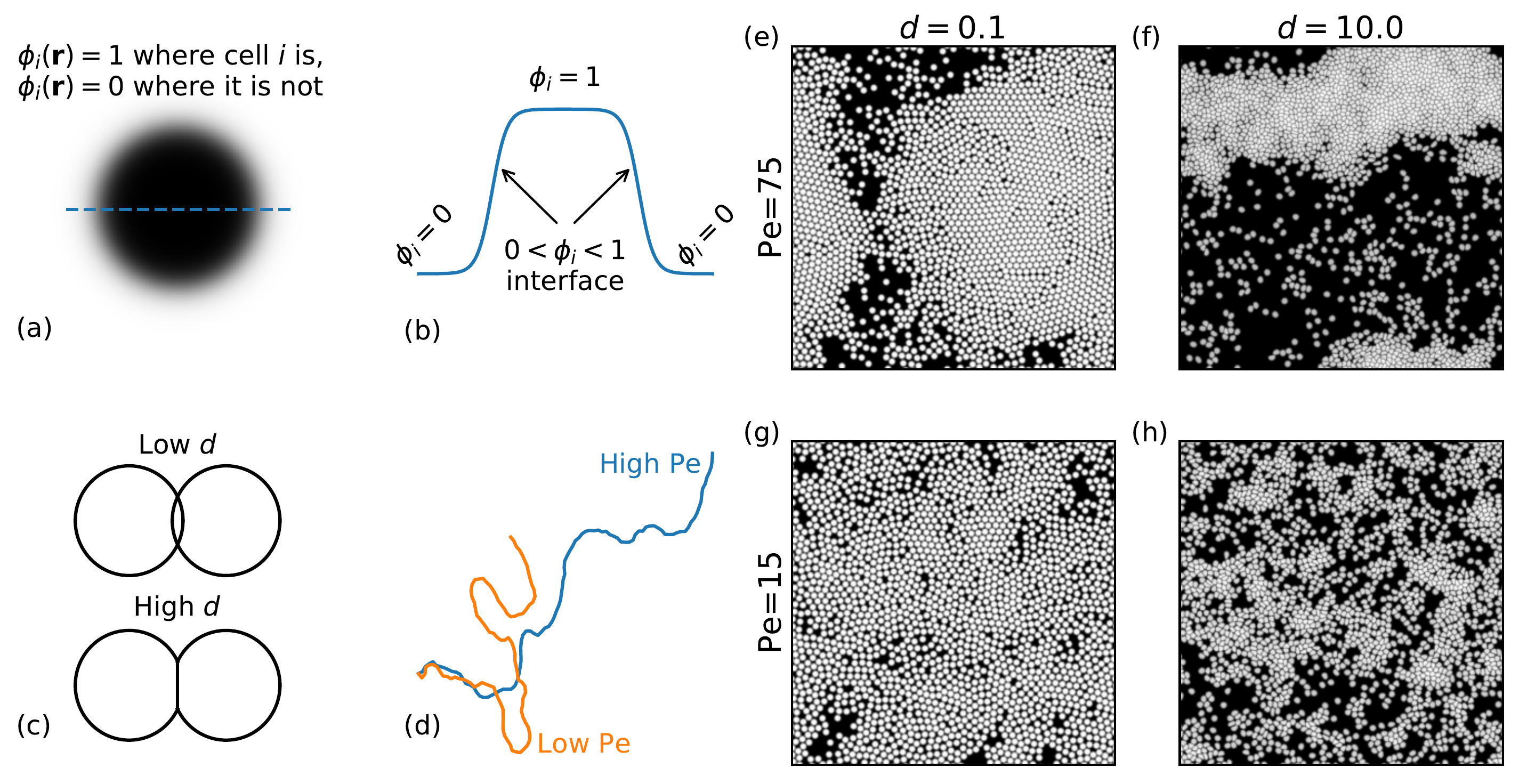}
 \caption{Illustrations of the model features (a)-(d) along with snapshots of the particles in the full simulations (e)-(h). (a) shows an isolated cell, which is circular. (b) shows a typical profile for the phase field of a cell. 
 \mc{(c) depicts how cells at a low $d$ will tend to overlap while maintaining a circular shape, while at a high $d$ cells will change shape to avoid overlapping. (d) illustrates representative cell trajectories at a lower and higher Pe.}
 The snapshots in (e)-(h) show the full system at low and high $d$ and $\text{Pe}$ as indicated.}
 \label{fig:model}
\end{figure*}

We model $N$ cells as deformable particles, each described by a phase field $\phi_i(\mathbf{r})$ \cite{nonomura2012study, lober2015collisions, palmieri2015multiple, mueller2019emergence, loewe2020solid, zhang2020active, balasubramaniam2021investigating, hopkins2022local, ritter2022differential, armengol2022epithelia}. The phase field model allows us to describe arbitrary cell shapes and to vary the cell edge tension. Phase field models have \mcm{been shown to capture}  many mechanical properties of tissue monolayers \cite{palmieri2015multiple,mueller2019emergence,armengol2022epithelia,ritter2022differential, balasubramaniam2021investigating}. The free energy of the system is 
\begin{equation}
\begin{split}
    \mathcal{F} = \sum_{i=1}^{N} \Bigg[ & \kappa \int d^2 \mathbf{r}\left( \phi_i^2 \left(\phi_i - 1 \right)^2 + \xi_0^2 \left( \mathbf{\nabla} \phi_i \right)^2 \right) \\ +
    & \lambda \pi R^2 \left(1 - \int d^2 \mathbf{r} \frac{\phi_i^2}{\pi R^2} \right)^2 + \epsilon \sum_{i<j=1}^N \int d^2 \mathbf{r}\, \phi_i^2 \phi_j^2 \Bigg]\;.
\end{split}
\end{equation}
The first term sets the field $\phi_i$ to be 1 in the interior of the cell and 0 in the exterior. The second term penalizes gradients in the field with a stiffness \mc{proportional to $\kappa$.} 
The third term sets the preferred area to that of a circle of radius $R=8$. An isolated cell will be circular, as in \fig \ref{fig:model} (a). The resulting $\phi_i$ profile will interpolate from 0 outside the cell to 1 inside the cell, over an interfacial thickness $\xi=2\xi_0$, which we keep fixed with $\xi_0=1$. The last term incorporates steric repulsion by penalizing overlap between different cells with strength $\epsilon=0.01$. 
When two cells interact, they may overlap or they may change their shape to avoid overlapping. We define the deformability $d$ as the ratio of the characteristic energy of overlap to the characteristic energy of shape deformation: $d = \frac{1}{24} (\epsilon \xi R)/(\sigma R) = \frac{\epsilon}{4 \kappa}$.
$\sigma=\kappa \xi_0/3$ is the cell-edge tension and the factor of $\frac{1}{24}$ brings this definition in line with previous work \cite{loewe2020solid}, in which $d \sim 1$ was shown to result in a qualitative change in the cellular interactions. Figure \ref{fig:model} (c) qualitatively illustrates how varying $d$ changes the interactions between cells. We also define the \mcm{cell} compressibility $\chi=\lambda/\epsilon$, which we keep fixed at $\chi=50$. This value of the compressibility allows for polydispersity while preventing cells from collapsing.

We model cells crawling on a substrate, which leads to the following evolution equation for the fields
\begin{equation}
    \frac{\partial \phi_i}{\partial t} + \mathbf{v}_i \cdot \mathbf{\nabla} \phi_i = -\frac{1}{\gamma} \frac{\delta \mathcal{F}}{\delta \phi_i}\;,
\end{equation}
where $\gamma$ = 1 is the inverse mobility. We incorporate the cell motility via the advection term in the field equation. The advection velocity is determined by self-propulsion and interaction terms arising from passive forces
\begin{equation}
    \mathbf{v}_i = v_0 \hat{\mathbf{p}}_i + \frac{1}{\Gamma A_i}\mathbf{f}_i\;,
\label{eq:advec}
\end{equation}
where \mcm{$\hat{\mathbf{p}}_i=\left( \cos\theta_i, \sin\theta_i \right)$ is the cell polarity which  determines the direction of isolated cell motion.}
The passive forces \mcm{are given by}
\begin{equation}
\mathbf{f}_i=-\sum_j^N \int d^2 \mathbf{r} \phi_i \phi_j \mathbf{\nabla} \mu_j\;,
\label{eq:force}
\end{equation}
with $\mu_j = \frac{\delta \mathcal{F}}{\delta \phi_j}$  the chemical potential of cell $j$, $A_j=\int d^2 \mathbf{r}\, \phi_j^2$ the cell's area, and $\Gamma$  a friction per unit area. We assume that all cells have the same self-propulsion  speed $v_0=0.0035$. The direction $\theta_i$ diffuses at a rate $D_r$, i.e., $d\theta_i(t) = \sqrt{2 D_r} dW_i(t)$, where $dW_i(t)$ is a Wiener process. We quantify the activity via the P\'eclet number $\text{Pe} = v_0/(R D_r)$, which is the ratio of the cell's persistence length $\ell_p=v_0/D_r$ to its size. The effect of varying $\text{Pe}$ on cell trajectories is illustrated in \fig \ref{fig:model} (d).

\section{Phase separation of deformable particles}
\label{sec:results}
To study motility-induced phase separation, we simulate $2,000$ cells in a square simulation box of length $L=897$, giving a packing fraction $\varphi=\frac{N \pi R^2}{L^2} = 0.5$.
At fixed packing fraction, we vary both $d$ and $\text{Pe}$.
When varying $\text{Pe}$, we fix $v_0=0.0035$ and vary $D_r$ (see Table \ref{tbl:parameters}). 
Since we work at a fixed interfacial thickness $\xi$, we vary $d$ by changing the cell edge tension $\sigma$.
We show snapshots from several of these simulations in \fig \ref{fig:model} (e)-(h).
At low $\text{Pe}$ the system does not phase separate regardless of the value of $d$.  \fig \ref{fig:model} (g,h). 
As $\text{Pe}$ is increased, phase separation occurs for both values of $d$ \fig \ref{fig:model} (e,f). 
One qualitative difference is that the dense phase in the high $d$ system takes up less area because the deformable cells can pack closely, while the rigid disk-like cells in the low $d$ system cannot pack as tightly.

\subsection{Deformability enhances phase separation}
\begin{figure}[h]
\centering
  \includegraphics[width=\linewidth]{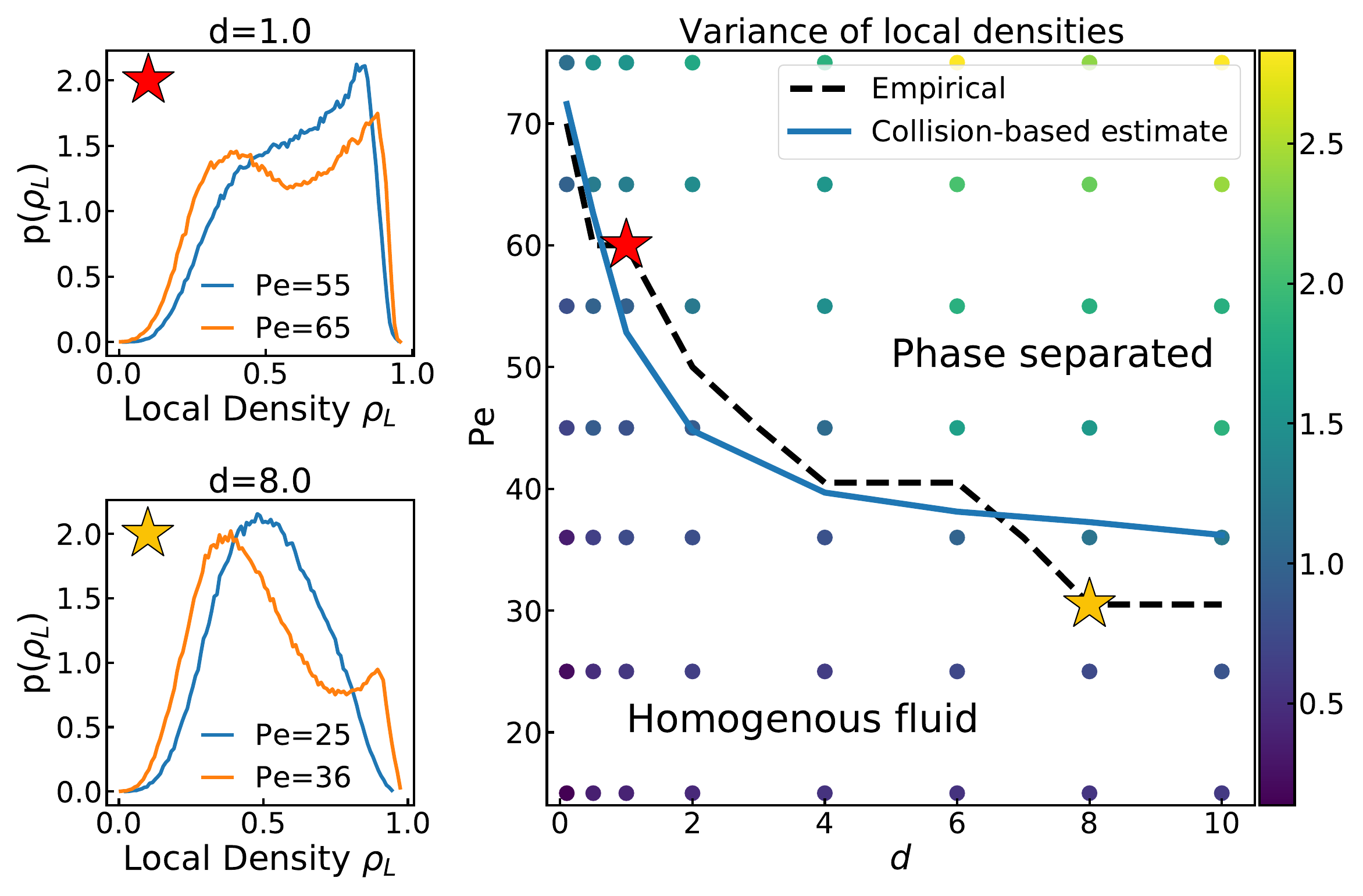}
  \caption{Left: local density curves for two different $d$, showing the difference in the distribution below and above the empirically determined cutoff. Right: phase diagram based on the variance of the local density distributions, with a fit based on the deformability dependence of the duration of two-body collisions. The stars indicate the deformabilities corresponding to the distributions on the left. The variance is normalized by the empirically chosen cutoff for phase separation.}
  \label{fig:pd}
\end{figure}

We find that, like in studies on rigid particles, phase separation  occurs above a critical $\text{Pe}$.
\mcm{To quantify the onset of phase separation we divide the system into square subsystems of size $10R$ } and calculate the local density $\rho_L$, defined as the area fraction of the subsystem where the local phase field is greater than 0.5. Examples of the distributions of $\rho_L$ are shown on the left side of \fig \ref{fig:pd}. 
This particular definition of the local density captures the fraction of the area that is excluded by steric interactions.
It also allows the most accurate comparison between low and high $d$ systems because the values are restricted between 0 and 1, even if the fields in tightly compressed cells reach values above 1.
We have verified that varying the subsystem size does not qualitatively change the distributions, as long as the subsystem size remains significantly larger than $R$ but smaller than $L$.
When the system phase-separates, the local density distribution changes from unimodal to bimodal. To quantify this change we use the variance of the distribution and choose a cutoff of 0.0378  to identify phase separation.
The resulting phase diagram is shown on the right side of \fig \ref{fig:pd}, where the variances have been normalized by this cutoff.
As one can see from the histograms on the left side of \fig \ref{fig:pd}, the chosen cutoff separates the homogeneous and phase separated systems. 
The required $\text{Pe}$ for phase separation decreases with increasing $d$.

\subsection{Deformability modifies the effective duration of a collision}
Previous work on motility-induced phase separation has shown that the transition from a homogeneous fluid to a phase separated state can be captured by continuum models formulated in terms of coarsed grained density  and polarity fields. \cite{fily2012athermal,stenhammar2013continuum,bialke2013microscopic,fily2014freezing}
Interactions renormalize the self-propulsion speed $v_0$, which becomes $v(\rho)$, a function of the density.
To capture the deformability dependence of the phase diagram, we start from the observation that at long times an isolated ABP behaves as a random walk of step length $\ell_p \simeq v_0 \tau_p$, with $\tau_p=D_r^{-1}$. Following the argument given in Ref.~\cite{stenhammar2013continuum}, we note that an ABP will be slowed down by collisions during each step of length $\ell_p$, resulting in a reduction of the effective step length $\ell(\rho) < \ell_p$.
Denoting by $n_c$ the number of collisions in $\tau_p$, and by $\tau_s$ the typical stalling time associated with each collision, the effective step length can be written as $\ell(\rho)=v_0(\tau_p - n_c \tau_s)$. The effective self-propulsion speed  is then given by $v(\rho)=\ell(\rho)/\tau_p$.
We estimate the number of collisions \bll{in} a time $\tau_p$  as 
$n_c\simeq \tau_p/\tau_{\text{mft}}$, where $\tau_{\text{mft}}$ is the mean free time between collisions. This is controlled by the scattering cross section and for circular particles can be written in terms of the number density $\rho$ as $\tau_{\text{mft}}=1/(2Rv_0\rho)$.
The effective propulsive speed can then be written as~\cite{fily2014freezing} 
\begin{equation}
    v(\rho) \simeq v_0 \left( 1 - \frac{\tau_s}{\tau_{\text{mft}}}\right)\;.
    \label{eq:vrho}
\end{equation}
This derivation makes sense at low densities, where $\tau_{\rm mft}\gg \tau_s$, and two body collisions are the primary cause of velocity slow-down. Eq.~(\ref{eq:vrho}) predicts a linear dependence of velocity on density. This has been observed empirically to hold to a good approximation up to much larger densities which might be {\it a priori} expected on the basis of this simple derivation~\cite{stenhammar2013continuum}.

Previous work \cite{fily2012athermal,stenhammar2013continuum,bialke2013microscopic,fily2014freezing} on continuum models of MIPS has shown that the onset of phase separation can be understood qualitatively by a linear instability associated with the vanishing of an effective diffusion coefficient, given by~\cite{fily2014freezing}
\begin{equation}
    \mathcal{D}\mcm{(\rho)} = \frac{v^2(\rho)}{2 D_r} \left(1+\frac{d\ln v(\rho)}{d\ln\rho}\right)\;.
    \label{eq:D}
\end{equation}
Within our approximation in Eq.~(\ref{eq:vrho}), MIPS ensues when $\tau_s\sim \tau_{\rm mft}$, which is also when such an approximation breaks down, suggesting that many-body collisions become important. 

\begin{figure}[t]
\centering
  \includegraphics[width=\linewidth]{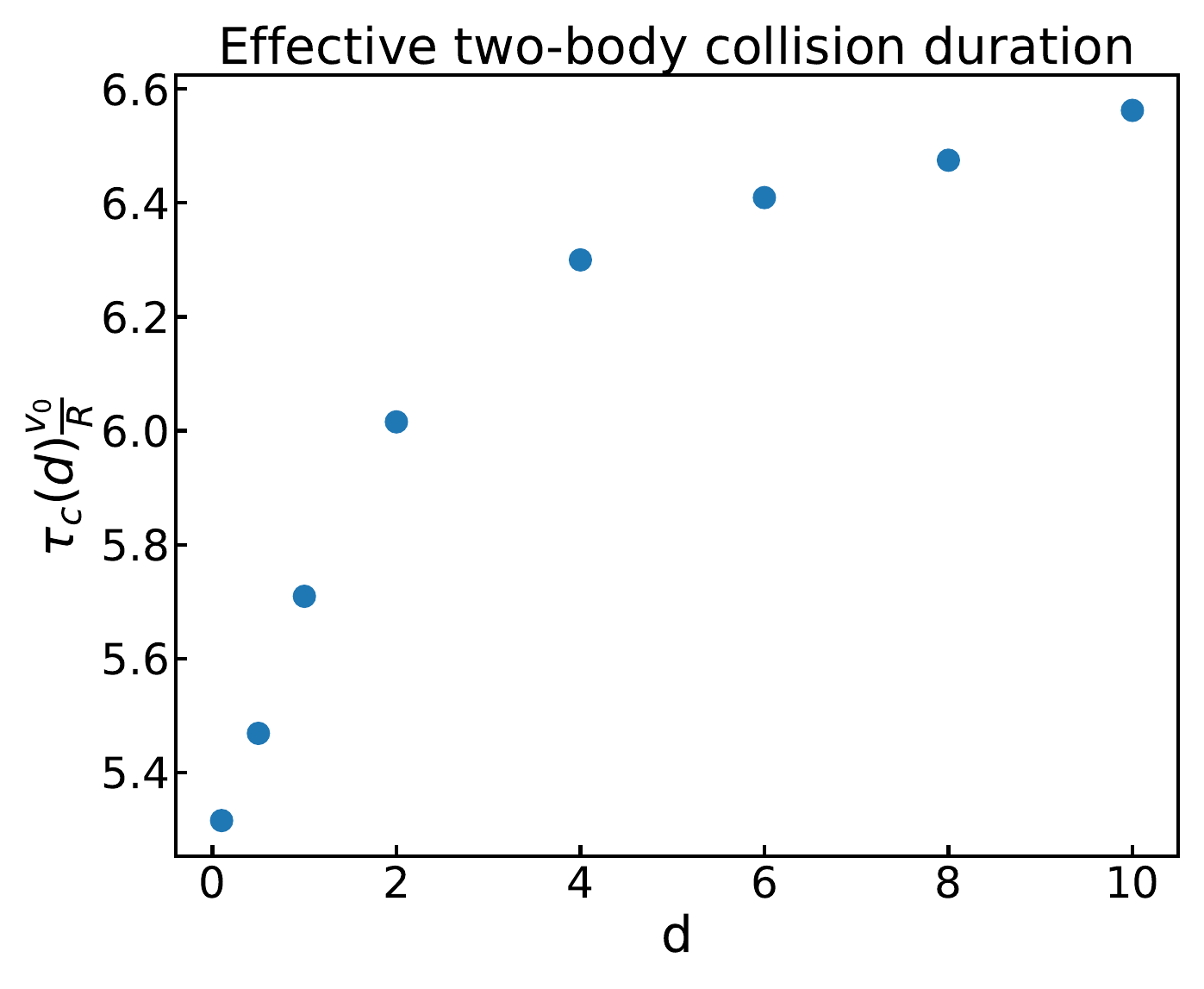}
  \caption{The collision duration $\tau_c$ as measured from two body simulations of nearly head-on collisions with $D_r=0$ increases with increasing particle deformability. See \ref{section:two-body} for more details. 
  }
  \label{fig:coll}
\end{figure}

To estimate the stalling time $\tau_s$, and hence the onset of MIPS in our deformable droplet system, we note that the latter is mainly controlled by two processes: the reorientation that occurs at rate $D_r$ and the fact that collisions among deformable particle have a finite duration $\tau_c$.  Assuming that $\tau_s$ is controlled by the faster of these two processes, we write $\tau_s^{-1}=a_1 D_r+a_2\tau_c^{-1}$, with $a_1$ and $a_2$ fitting parameters expected to be of order one. We then estimate the deformability dependence of \bll{$\tau_c$, i.e., the time it takes for particles to move past one another due to interactions,} by examining numerically two body simulations of nearly head-on collisions in the limit $D_r=0$, where particles cannot escape the collision by turning their nose. We find that $\tau_c$ depends strongly on deformability (\fig \ref{fig:coll}).  
Using this estimate in 
the instability condition 
$\tau_{\text{mft}}(\rho)=\tau_s(d)$
we obtain the dashed line in \fig \ref{fig:pd}, with $a_1=3.70$ and $a_2=1.42$. 
Therefore we find that the criterion $\tau_s=\tau_{\rm mft}$ predicts well the onset of MIPS, so that the strong dependence of the stalling time on deformability  obtained from two body collisions captures the increasing propensity of more deformable particles to phase separate.

\subsection{Structure of the dense phase}

\begin{figure}[h]
\centering
  \includegraphics[width=\linewidth]{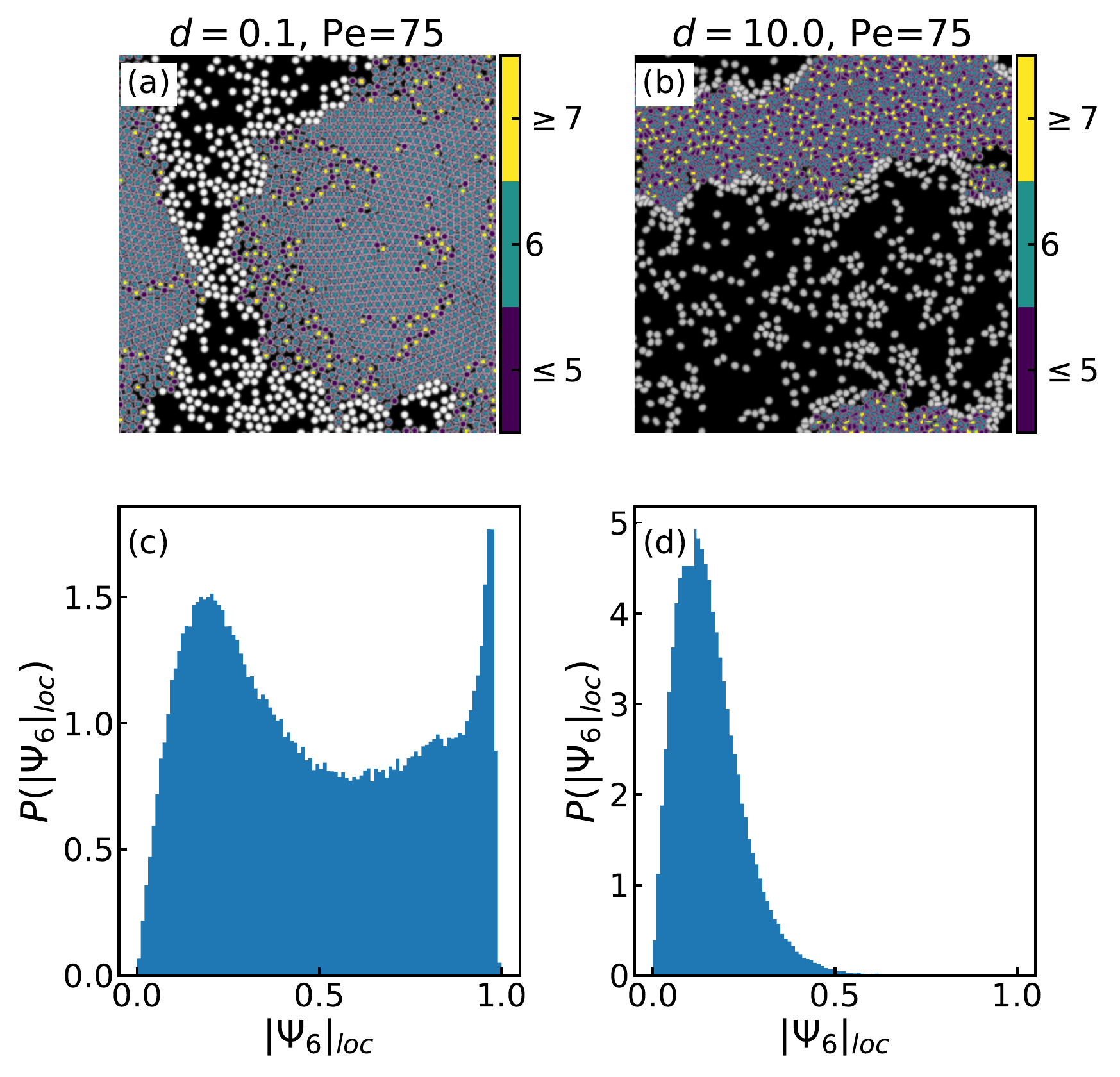}
  \caption{Top row:  snapshots of the phase-separated system with some of the particles in the dense phase colored by the number of neighbors as determined by a Voronoi diagram of the centers of mass for low \mc{((a), $d=0.1$) and high ((b), $d=10$)} deformability. Bottom row: corresponding (time-averaged) distributions of the local bond-orientational order parameter $|\Phi_6|_{loc}$.}
  \label{fig:bond_order}
\end{figure}

The  dense phase  becomes more disordered as deformability increases.
\mcm{This is evident} from \fig \ref{fig:bond_order}(a) and (b), where we show snapshots in which the cells are colored by their number of neighbors as determined by a Voronoi construction of their centers of mass. Clearly the number of structural defects increases with deformability.
To quantify the structure, we define the bond-orientational order of a cell $j$ as $\Psi_{6,j}=\frac{1}{N_{nn,i}}\sum\limits_{k \in nn} e^{i6\theta_{j,k}},$ where the sum is over the $k$ nearest neighbors and $\theta_{j,k}$ is the angle between the center of mass of cell $j$ and cell $k$.
We define the local bond-orientational order $|\Psi_6|_{\text{loc}}$ as the average of $\Psi_{6,j}$ over the cells within a subsystem of size $10R$.
We show the distributions of this quantity in \fig \ref{fig:bond_order}, for low (c) and high (d) deformability. 
At low deformability the distribution has a peak near $|\Psi_6|_{\text{loc}}\sim 1$, corresponding to local hexatic order of the dense clusters, and a second peak at a small values of $|\Psi_6|_{\text{loc}}$ arising from the disordered low density gas. 
For high deformability, however, there is no local heaxatic order in the dense clusters, and the distribution has a single peak at low $|\Psi_6|_{\text{loc}}$. 
The increase in disorder as a function of deformability is similar to the change seen in the confluent deformable particle monolayer studied in previous work. \cite{hopkins2022local}


\section{Summary}
\label{sec:summary}
In summary, we have characterized numerically the phase diagram of a system of purely repulsive deformable active particles as a function of their deformability and motility. This case is an important one to consider when the applicability of motility-induced phase separation to biological systems, such as cell suspensions, is considered: indeed, cells behave differently from colloidal rigid particles, and can be better represented by deformable droplets.

We have shown that, like rigid APBs, deformable particles phase separate into a dilute and dense phase for sufficiently persistent motility. However, we found that deformability has two important effects on motility-induced phase separation. First, deformable particles are able to phase separate at a significantly lower motility than rigid ones. This effect can be explained by the fact that deformability increases the duration of two-body collisions, thereby enhancing the slow-down of motility induced by crowding. Second, deformability strongly affects the nature of the high-density phase, which is glassy for squishy (more deformable) particles, which become polydisperse, and near-crystalline for rigid (less deformable) particles. It would be of interest to study in the future the dynamics within the high-density phase, to assess whether the structural differences we have observed translate into a dynamical phase transition between the two regimes.

We stress that increasing deformability in our work has a distinctly different effect from the softening of the repulsive interaction.
Previous work on simulations of rigid repulsive ABPs has shown that softening the repulsive interaction suppresses both motility-induced phase separation and bond orientational order \cite{sanoria2021influence}.
Phase separation in that context is suppressed because softer repulsive interactions allow particles to overlap, which reduces the amount that particles are slowed down due to collisions.
In contrast, deformability, as implemented in our work, suppresses overlap and enhances the slow down due to collisions, which promotes phase separation.

Future work will be needed to further connect MIPS to biological systems.
Additional interactions beyond steric repulsion, such as differential adhesion \cite{foty2005differential}, as well as chemically mediated interactions~\cite{liebchen2017phoretic}, may enhance cell aggregation or affect pattern formation in real systems.
A further interesting generalization would be to consider mixtures of deformable and rigid particles, which could lead to sorting between cells within the high-density phase. 

\section*{Author Contributions}
Conceptualization: A.H., B.L. and M.C.M.; Data curation: A.H. and B.L.; Formal analysis: A.H. and B.L.; Funding acquisition: M.C.M. and D.M.; Investigation: A.H. and B.L.; Methodology: all; Software: A.H., B.L., and M.C.; Supervision: M.C.M.; Visualization: A.H. and B.L.; Writing - original draft: A.H.; Writing - review \& editing: all

\section*{Conflicts of interest}
There are no conflicts to declare.

\setcounter{section}{0}
\renewcommand{\thesection}{Appendix \Alph{section}}
\setcounter{equation}{0}
\renewcommand{\theequation}{A\arabic{equation}}
\section{Role of interaction forces in the advection equation}
\begin{figure}

 \includegraphics[width=\linewidth]{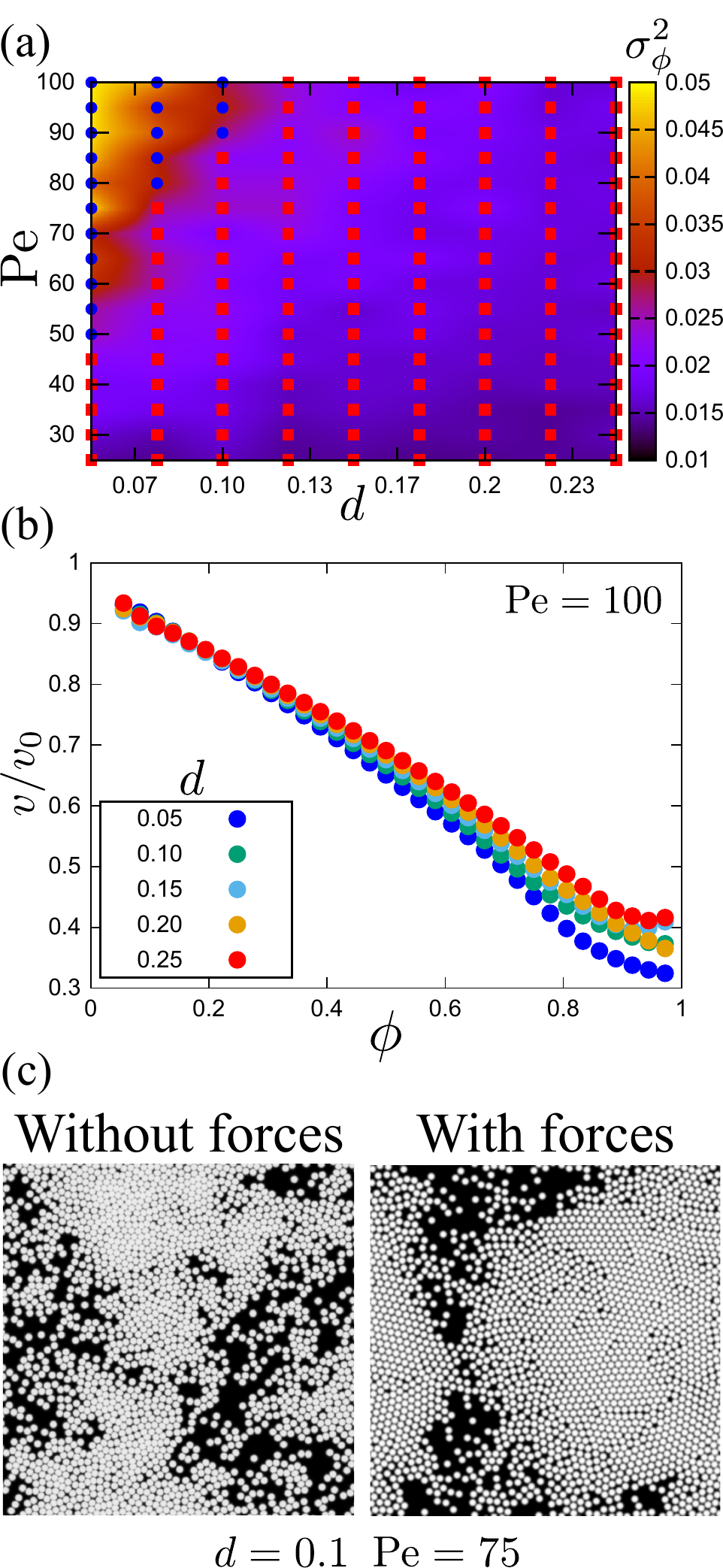}
 \caption{(a) Phase diagram \mcm{obtained} simulations \mcm{using the model of Ref. \cite{loewe2020solid}} without passive forces in Eq.~(\ref{eq:advec}). Colorbar: variance of the local packing fraction. Blue dots: Phase-separated states (bimodal distribution of packing fractions). Red dots: Homogeneous state (unimodal distribution of packing fractions).
 In the absence of passive forces, deformability suppresses phase separation. (b) Cell velocity as a function of local packing fraction for cells without passive forces. Increasing deformability reduces the slow-down induced by \mcm{crowding}, thus \mcm{suppressing} MIPS.
 (c) \mcm{Snapshot of phase separated states} with and without passive forces at the same deformability and motility \bl{($d = 0.1$, $\text{Pe} = 75$).} Passive forces lead to a more spread-out spatial distribution of cells.
 \label{fig:appendix}}
\end{figure}

\mcm{Different phase field models have been used in previous the literature that have included \bl{\cite{palmieri2015multiple,mueller2019emergence}} or neglected \bl{\cite{loewe2020solid, armengol2022epithelia}} explicit}
interaction forces in \eq \ref{eq:advec} for the cell advection velocity. 
We \mcm{show here that the presence/absence of} 
these forces \mcm{has a} significant \mcm{effect on MIPS.}
Notably, the \mcm{effect} of deformability \mcm{on} MIPS is reversed, as \mcm{when passive interaction forces are not included in the force balance equation for the advection velocity,} increasing deformability suppresses, \mcm{instead of enhancing,} phase separation (Fig.~\ref{fig:appendix}(a)). This is a strikingly strong effect, as just small increases in deformability destroy the phase-separated state, even at \mcm{high} motility (Fig.~\ref{fig:appendix}(a)). \mcm{This is perhaps not surprisingly, given that the density-dependent suppression of propulsive speed that is the hallmark of MIPS comes precisely from the component of the mean repulsive force along the direction of each particle polarization~\cite{bialke2013microscopic}.}  This reversal in the role of deformability suggests that deformable cells behave as soft disks in the absence of passive forces. Increasing deformability is akin to increasing the softness of the disks, thus hampering phase separation by virtue of colliding cells passing  more easily through each other. This effect can also be appreciated by measuring the cell speed as a function of the local packing fraction. In general, increasing the local packing fraction slows down the cells. \mcm{In the absence of passive forces,} however, increasing deformability diminishes this effect (Fig.~(\ref{fig:appendix}(b)), leading to an increased speed in crowded environments,  suggesting that cells can more easily squeeze through their neighbours. Finally, this effect can also be observed by visually inspecting the \mcm{phase separated} system (see \ref{fig:appendix}(c)): passive forces help cells keep away from each other, leading to  larger clusters. In their absence, the same deformability and motility lead \mcm{to sparser clusters.}

\section{Details of two-body collisions}
\label{section:two-body}
\begin{figure}[h]
\centering
  \includegraphics[width=\linewidth]{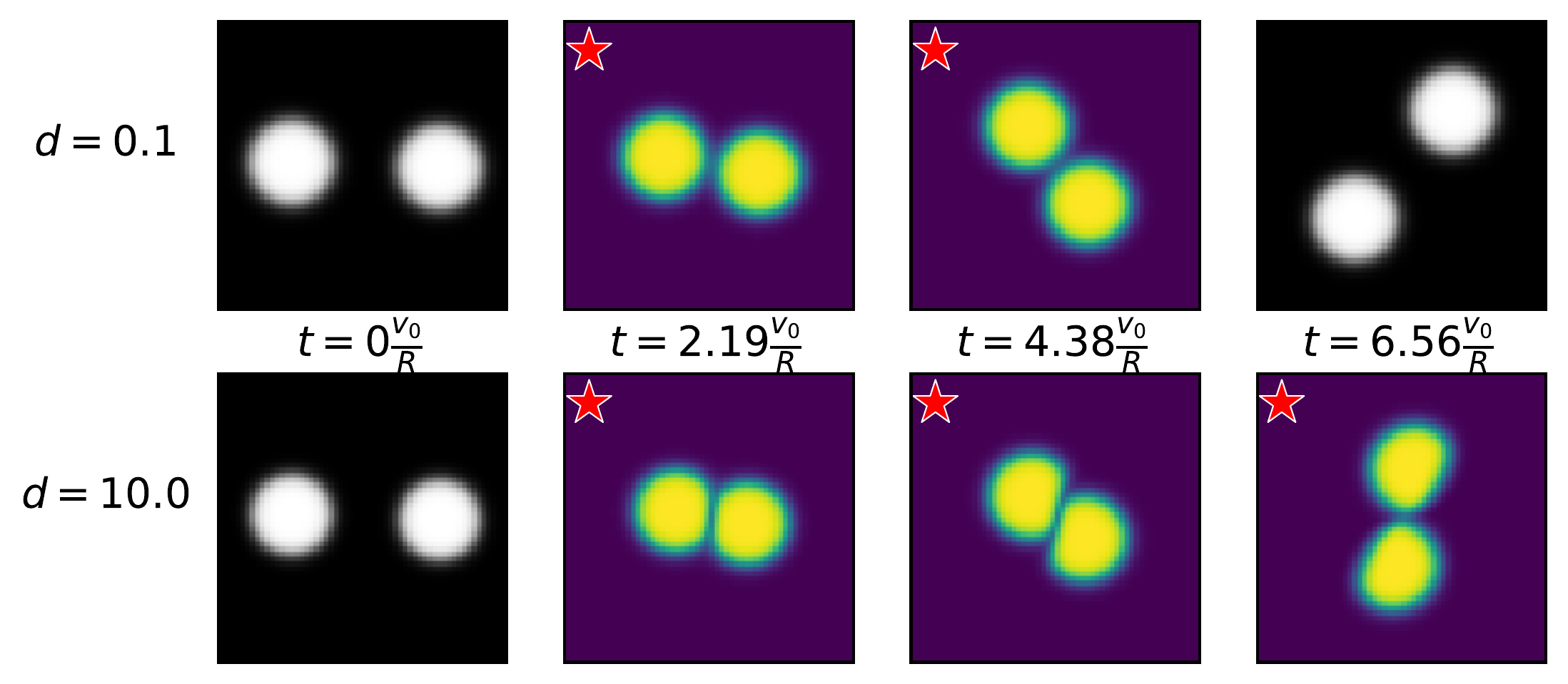}
  \caption{Snapshots of two-body collisions. Frames in which the cells are not in contact are colored in black and white. When the cells are in contact, the frames are colored in purple and yellow and are also indicated by a star in the top left corner.}
  \label{fig:coll_snaps}
\end{figure}

We consider two particles, initially isolated and circular, which propel \mcm{towards each other} in a head-on collision.
All the cell parameters are as in the main text, except $D_r=0$ to eliminate rotational noise.
We offset the particles by one lattice point (impact parameter $b=\frac{1}{8}R$) so that they are nearly head-on, but are still able to move past one another \mcm{in the absence of}  noise.
We focus on nearly head-on collisions because those are the ones which significantly slow down a particle, and hence are the most important for cluster formation.
We choose a cutoff of $\phi_1+\phi_2=0.1$ to define when the two particles are in contact with one another, and measure the collision duration as the total time the particles are in contact.
As can be seen in \fig \ref{fig:coll_snaps}, when particles with high $d$ collide \mcm{(bottom row)}, their shape changes, which slows their motion past one another.

\section{Table of Parameters}
\begin{table}[H]
\small
  \caption{\ Value(s) of the parameters used in the simulations.}
  \label{tbl:parameters}
  \begin{tabular*}{0.48\textwidth}{@{\extracolsep{\fill}}llll}
    \hline
    Parameter & Interpretation & Dimensions & Value(s) \\
    \hline
    $d$ & deformability & - & 0.1 -- 10 \\
    $R$ & cell radius & [L] & 8 \\
    $\xi$ & cell interface thickness & [L] & 2 \\
    $\epsilon$ & strength of repulsion & [E] [L]$^{-2}$ & 0.01 \\
    $\chi$ & cell compressibility & - & 50 \\
    $\gamma$ & inverse mobility & [E][T][L]$^{-2}$ & 1 \\
    $\Gamma$ & substrate friction density & [E][T][L]$^{-4}$ & 1/64 \\
    $v_0$ & cell self-propulsion speed & [L][T]$^{-1}$ & 0.0035 \\
    $D_r$ & polarity diffusion rate & [T]$^{-1}$ & $5.8\times 10^{-6}$ \\ &&& -- $2.9\times 10^{-5}$ \\
    $dt$ & time step & [T] & 0.5 \\
    $\varphi$ & packing fraction & - & 0.5 \\
    
    \hline
  \end{tabular*}
\end{table}

\section*{Acknowledgements}
The work by A.H. and M.C.M. was supported by the National Science Foundation Grant No. DMR-2041459. \bl{This research has received funding (B. L.) from the European Research Council under the European Union’s Horizon 2020 research and innovation programme (Grant Agreement No. 851196).} Use was made of computational facilities purchased with funds from the National Science Foundation (CNS-1725797) and administered by the Center for Scientific Computing (CSC). The CSC is supported by the California NanoSystems Institute and the Materials Research Science and Engineering Center (MRSEC; NSF DMR 2308708) at UC Santa Barbara.



\balance


\bibliography{mips} 
\bibliographystyle{rsc} 

\end{document}